\documentclass[twocolumn,floatfix,superscriptaddress,longbibliography,aps,prl]{revtex4-1}

\usepackage{amsmath}
\usepackage{amssymb}
\usepackage{gensymb}
\usepackage{bm,physics}
\usepackage{graphicx}
\usepackage{float}
\usepackage{placeins}
\usepackage{braket}
\usepackage{bbold}
\usepackage{ulem}
\usepackage[colorlinks, linkcolor=blue, citecolor=blue, urlcolor=blue, breaklinks=red]{hyperref}

\makeatletter
\newcommand{\mypm}{\mathbin{\mathpalette\@mypm\relax}}
\newcommand{\@mypm}[2]{\ooalign{%
  \raisebox{.1\height}{$#1+$}\cr
  \smash{\raisebox{-.6\height}{$#1-$}}\cr}}
\makeatother

\begin{document}

\title{Subradiance with saturated atoms: population enhancement of the long-lived states}

\author{A. Cipris}
\affiliation{Universit\'e C\^ote d'Azur, CNRS, Institut de Physique de Nice, France}
\author{N. A. Moreira}
\affiliation{Instituto de F\'isica de S\~{a}o Carlos, Universidade de S\~{a}o Paulo - 13566-590 S\~{a}o Carlos, SP, Brazil}
\author{T. S. do  Espirito  Santo}
\affiliation{Instituto de F\'isica de S\~{a}o Carlos, Universidade de S\~{a}o Paulo - 13566-590 S\~{a}o Carlos, SP, Brazil}
\author{P. Weiss}
\affiliation{Universit\'e C\^ote d'Azur, CNRS, Institut de Physique de Nice, France}
\author{C. J. Villas-Boas}
\affiliation{Departamento de F\'isica, Universidade Federal de S\~{a}o Carlos, 
Rod.~Washington Lu\'is, km 235 - SP-310, 13565-905 S\~{a}o Carlos, SP, Brazil}
\author{R. Kaiser}
\affiliation{Universit\'e C\^ote d'Azur, CNRS, Institut de Physique de Nice, France}
\author{W. Guerin}
\affiliation{Universit\'e C\^ote d'Azur, CNRS, Institut de Physique de Nice, France}
\author{R. Bachelard}
\affiliation{Departamento de F\'isica, Universidade Federal de S\~{a}o Carlos, 
Rod.~Washington Lu\'is, km 235 - SP-310, 13565-905 S\~{a}o Carlos, SP, Brazil}

\begin{abstract}
Dipole-dipole interactions are at the origin of long-lived collective atomic states, often called subradiant, which are explored for their potential use in novel photonic devices or in quantum protocols. Here, we study subradiance beyond linear optics and experimentally demonstrate a two hundred-fold increase in the population of these modes, as the saturation parameter of the driving field is increased. We attribute this enhancement to a mechanism similar to optical pumping through the well-coupled superradiant states. The lifetimes are unaffected by the pump strength, as the system is ultimately driven toward the single-excitation sector.
\end{abstract}

\date{\today}

\maketitle

Light is an excellent tool to encode and transmit information, yet it comes up short in terms of storage. It is then convenient to `write' the information into a material memory, before `reading' it out at a later time. Atoms and their artificial versions are natural candidates to fulfill that purpose, where photons are converted into atomic excitations. In this context, cold atoms benefit a substantial cross-section to couple to light, and provide access to a broad range of lifetimes, with transitions linewidths ranging from mHz to MHz, making them useful tools for quantum information processing~\cite{Hammerer:2010} and quantum metrology~\cite{Pezze2018}, for instance.

Considering interactions between the atoms opens yet new possibilities to harness their potential. In particular, the dipole-dipole interaction, which rises precisely in presence of photons, leads to a variety of collective responses \cite{Kupriyanov2017,Guerin2017b}, such as single-layer atomic mirrors~\cite{Rui2020}, superradiance~\cite{Araujo2016, Roof2016, Ortiz2018, Okaba2019} and subradiance~\cite{Bienaime2012,Guerin2016,Solano2017,Das2020}. The two latter effects correspond to modes of lifetimes orders of magnitude smaller or larger, respectively, than the single-atom one, and the use of external fields has been proposed to transfer excitations between modes and thus realize write and read operations~\cite{Scully2015,Facchinetti2016}. Platforms such as atoms coupled to fibers~\cite{Sprague2014,Sayrin2015,Gouraud2015,Cho2016} and superconducting qubits~\cite{Wang2020} have also demonstrated their potential to manipulate such states.



These subradiant states are, by essence, difficult to drive, due to their weak coupling to the external world. The protocols designed to address them have been tailored for the single-excitation regime~\cite{Kalachev2006,Kalachev2007,Scully2015,Facchinetti2016}, which represent a drop in the sea of the long-lived states originally predicted by Dicke~\cite{Dicke1954}.

\begin{figure}[t]
\includegraphics[width=0.45\textwidth]{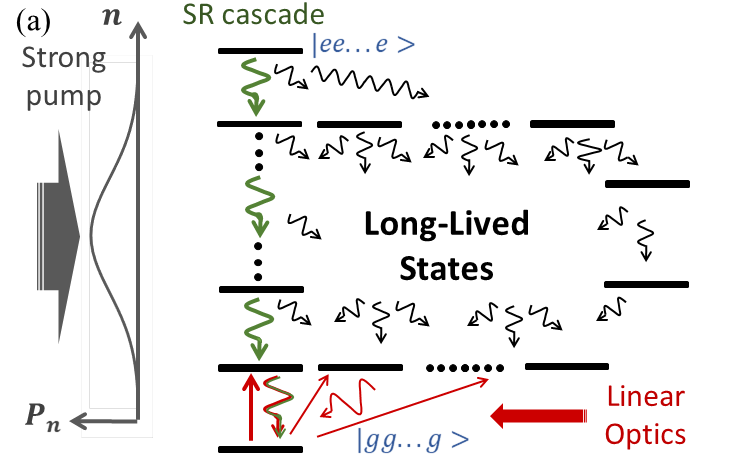} 
\vspace{0.25cm}
\includegraphics[width=0.45\textwidth]{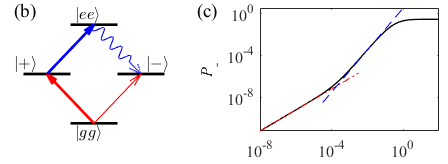} 
\caption{(a) Dicke space for $N\gg 1$ two-level atoms, where the downward arrows depict the decay processes. The processes in green refer to the Dicke superradiant cascade through symmetric states~\cite{Dicke1954}, and the red ones to the linear-optics (single-excitation) processes. $P_n$ schematically shows the population of the $n$-excitation states for a strong pump, illustrating the spread over various excitation numbers. (b) Energy levels for $N=2$ coupled atoms (see main text for details). (c) Computed population of the long-lived $\ket{-}$ state for $N=2$ coupled atoms. The red dash-dotted curve corresponds to $P_-=s_-/2\propto s$, while the dashed blue one scales as $s^2$. Simulation realized for $N=2$ atoms distant of $kr=0.05$, aligned with the pump axis, with a detuning $\Delta=-500\Gamma$.}\label{fig:scheme}
\end{figure}

In this work, we explore the many-excitation regime by increasing the pump strength and we report on a large increase of the excitations cast in the long-lived modes. This is interpreted as a process analogous to optical pumping [see Fig.\,\ref{fig:scheme}(a)]: addressing the multi-excitation superradiant states, well coupled to the external drive, allows one to efficiently populate the long-lived states through decay processes. Using numerical simulations, the study of the dynamics of the many-excitation states reveals that the longest lifetimes are found in the modes with fewer excitations, toward which the system quickly decays. 

The atomic cloud is modelled as an ensemble of $N$ two-level emitters with positions $\mathbf{r}_j$, a transition frequency $\omega_a=kc=2\pi c/\lambda$ between their ground and excited states $g$ and $e$ ($\sigma_m^-=\dyad{g_m}{e_m}$ and $\sigma_m^+=\dyad{e_m}{g_m}$ the loewring and rising operators), and a transition linewidth $\Gamma$. The cloud is driven by a near-resonance monochromatic field with Rabi frequency $\Omega(\mathbf{r})$, detuned from the transition by $\Delta$. The dipole-dipole interaction relies on the coupling of the atomic dipoles through common radiation modes, which results in sub- and superradiant collective modes (see \cite{SuppMat} and Refs.~\cite{Stephen1964, Lehmberg1970, Friedberg1973, Asenjo2017, Espirito2020} for details on the model).

Although our main focus is many-atom subradiance, let us first discuss the case of a pair of close atoms ($r_{12}\ll\lambda$) as a toy model, since it already captures the main features of our scheme. 
We consider a pump whose propagation axis is aligned with the two atoms ($\Omega(\mathbf{r})=\Omega_0 e^{ikz}$ and $\mathbf{r}_2-\mathbf{r}_1=r_{12}\hat{z}$), with frequency far from any resonances ($|\Delta|\gg\Gamma,\ \Delta_{12}$, with $\Delta_{12}$ the energy shift of the collective modes), to avoid specific effects such as blockade~\cite{Cidrim2020,Williamson2020} or anti-blockade~\cite{Almutairi2011} of excitations. The dipole-dipole interaction generates two collective single-excitation eigenstates $\ket{\pm}=(\ket{eg}\pm\ket{ge})/\sqrt{2}$, in addition to the ground $\ket{gg}$ and double-excited $\ket{ee}$ states [see Fig.\,\ref{fig:scheme}(b)]. These two collective states present an energy shift $\pm\Delta_{12}/2$, here neglected since $|\Delta|\gg\Delta_{12}$, and decay rates $\Gamma_{\mypm}=\Gamma\pm\Gamma_{12}$. Importantly, the pump couples mostly to the superradiant $\ket{+}$ state, and very weakly to the long-lived $\ket{-}$ one, and we introduce the effective Rabi frequency for each mode: $\Omega_+=\sqrt{2}\cos(kr_{12}/2)\Omega_0$ and $\Omega_-=\sqrt{2}\sin(kr_{12}/2)\Omega_0$ (up to a phase), which can be identified by rewriting the driving Hamiltonian in terms of the states $\ket{+}$ and $\ket{-}$.

The steady-state population of the long-lived mode then presents three typical regimes, depending on the pump strength. First, for the lowest intensities (linear-optics regime), the population of $\ket{ee}$ is negligible and the single-excitation modes $\ket{\pm}$ are driven only directly from the pump, so one obtains the following scaling for their population: $P_{\mypm}\approx s_{\mypm}/2\propto s$, with $s_{\mypm}=2\Omega_{\mypm}^2/(\Gamma_{\mypm}^2+4(\Delta\mp\Delta_{12}/2)^2)$ the effective saturation parameter for each mode, and $s=2\Omega_0^2/(\Gamma^2+4\Delta^2)$ the single-atom one [see Fig.\,\ref{fig:scheme}(c)].

As the drive strength is increased, the doubly-excited state $\ket{ee}$ is substantially populated thanks to the strong coupling of the drive to the superradiant state: $P_+\approx s_+/2\propto s$ and $P_{ee}\propto s^2$. Then, the $\ket{-}$ states gets an additional population by decay from $\ket{ee}$, at rate $\Gamma_-$, leading to a long-lived population that grows quadratically with the saturation parameter: $P_-\propto s^2$ [see Fig.\,\ref{fig:scheme}(c)].

Finally, for the largest values of the saturation parameter, i.e., with a Rabi frequency much larger than the involved interaction energy ($\Omega_0\gg\Delta\gg\Gamma,\ \Delta_{12}$), the population saturates as the system is cast into a separable state described by the density matrix $\hat{\rho}=\otimes_{j=1,2} (\ket{g_j}+\ket{e_j})(\bra{g_j}+\bra{e_j})/2$. This mixed state projects equally on the states $\ket{gg}$, $\ket{+}$, $\ket{-}$ and $\ket{ee}$, resulting in $P_-\approx1/4$. 
Hence, the strong pump overcomes the interaction energy that prevents, in the linear-optics regime, an efficient population of subradiant states.
From another perspective, the present mechanism is analogous to optical pumping, where an excited state (here $\ket{ee}$) is directly driven by the laser, and induces a population in the long-lived state (here $\ket{-}$) by incoherent decay.

Despite its extreme simplicity, let us now discuss how the $N=2$ case captures the essential features of our many-atom experiment, based on a cold atomic cloud of $N\approx 6\times10^9$ randomly distributed $^{87}$Rb atoms prepared in a magneto-optical trap. A detailed description of the setup and of the methods for observing subradiance can be found in \cite{Guerin2016,Weiss2018,Weiss2019}. In this new series of experiment, an extra care has been taken to control the possible detrimental effects of the large intensity probe on the atomic cloud: we reduced the pulse duration to $5\,\mu$s and added a repumper pulse between each probe pulse. Moreover, we used the improved characterization of the sample, as described in \cite{Weiss2019} and detailed in \cite{SuppMat}. We varied the saturation parameter in the range $3\times10^{-3}\lessapprox s(\Delta) \lessapprox2$ by varying the intensity of the probe beam \cite{SuppMat}.

\begin{figure}[t]
\includegraphics[width=0.45\textwidth]{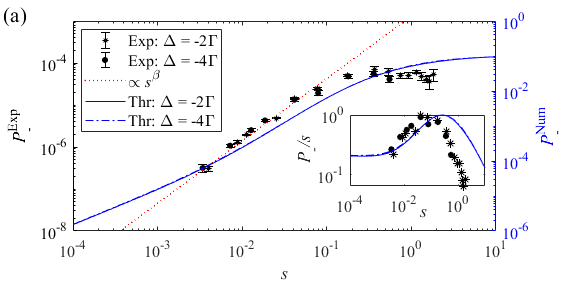} 
\includegraphics[width=0.45\textwidth]{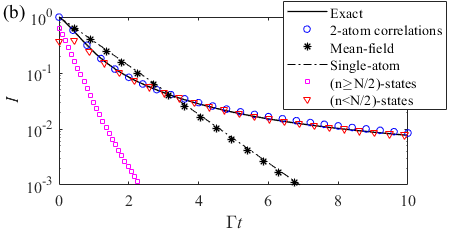} 
\caption{(a) Normalized population $P_-$ of long-lived states as a function of the saturation parameter. Experimental data (symbols, left axis) acquired for $b_0 =54\pm2$, with error bars describing the 95\% confidence bounds (statistical uncertainty only). Simulations (lines, right axis) realized for $b_0\approx 5$, averaging over $40$ realizations (error bars of order 1\%, not shown here). The red dotted line stands for a fit $P\propto s^\beta$, with $\beta\approx 1.49$. Inset: Ratio population over saturation parameter $P_-/s$, with a maximum normalized to unity. (b) Dynamics of the radiated intensity after switching-off the pump at $t=0$, for $N=7$ atoms, $b_0=5$ (using $60$ realizations), $\Delta=-2\Gamma$ and $s=50$. Simulations realized using exact simulations (plain curve), the pair-correlation approach (circles) and the mean-field model (stars). The contribution from states with $n\geq 4$ (squares) and $n\leq 3$ (triangles) are computed separately, for the exact simulations.\label{fig:PopuI}}
\end{figure}

To obtain the amplitude $A_-$ and the lifetime $\tau_-$ of the long-lived radiation we fit the collected intensity by an exponential $I(t)=A_-\exp(-t/\tau_-)$ in a range $t\in[150;250]/\Gamma$ (see \cite{SuppMat} for a few decay curves). The normalized population $P_- \propto A_- \tau_- /N$ of these long-lived states, defined as the number of excitations divided by the atom number, can be deduced by assuming that the long-lived excitations are radiated isotropically, and taking into account the collection efficiency of the detection. The measured population is presented in Fig.\,\ref{fig:PopuI}(a) for a fixed $b_0 = 54$. It undergoes a 200-fold increase, from $3\times10^{-7}$ to $7\times10^{-5}$, as the saturation parameter is increased from $s\approx 3\times10^{-3}$ to $0.3$. This corresponds to a maximum number of $\approx 4\times10^5$ excitations in these long-lived modes. 
Note that these numbers are only orders of magnitude since the detection efficiency (solid angle, quantum efficiency of the detector, various losses on the optical path) is not precisely calibrated. More importantly, we observe a super-linear scaling of the population, $P_-\propto s^\beta$ with $\beta\approx 1.49$, as illustrated in the inset of Fig.\,\ref{fig:PopuI} where $P_-/s$ is plotted, before the population saturates. We have checked that this scaling is reproduced  for several values of $b_0$~\cite{SuppMat}.

Furthermore, subradiant modes with different lifetimes can be accessed by considering different fitting windows: their population not only presents a similar behaviour (i.e., a super-linear growth as $\sim s^{1.5}$, and a maximum value of $\approx 10^{-4}$), but the curves also strongly overlap before they saturate \cite{SuppMat}. This suggests that despite longer-lived modes are expected to be less populated in a linear-optics scenario (because of the weaker coupling to the external world), the much larger connectivity between the modes introduced by the strong pump and the decay channels enhances their population. 

This super-linear growth of the long-lived population is a clear indication of beyond-linear-optics pumping of these modes and calls for a more specific study. Yet such a study requires addressing its size-$2^N$ Hilbert space, which unavoidably leads to drastic approximations. First, we note that the lifetime of collective modes in these clouds has been shown to scale with the on-resonance optical thickness $b_0=\sigma_\textrm{sc} \int \rho_a(0,0,z)dz$, both for superradiant~\cite{Araujo2016,Roof2016} and subradiant ones~\cite{Guerin2016} (with the cloud center and the pump waist at the origin, $\mathbf{k}_\mathrm{laser}=k\hat{z}$ and $\sigma_\textrm{sc}$ the atom cross-section). This allows one to study collective effects numerically by adjusting the resonant optical thickness of clouds of hundreds to thousands of particles, as available in simulations.

Second, simulating the dynamics of more than a dozen saturated two-level atoms requires additional approximations in describing the system state. We here resort to a truncated scheme based on the Bogoliubov--Born--Green--Kirkwood--Yvon approach, where the density matrix is recast as a sum of reduced density matrices of order $m=1..N$, thus establishing a hierarchy of quantum correlations~\cite{Bonitz2016}. 
The truncation of the hierarchy to two-particle quantum correlations has proven to be an efficient technique to simulate the dynamics of strongly driven atomic clouds~\cite{Kramer2015,Pucci2017}, and we refer to these references for further details. 

\begin{figure*}[!t]
\centering
\includegraphics[width=0.285\textwidth]{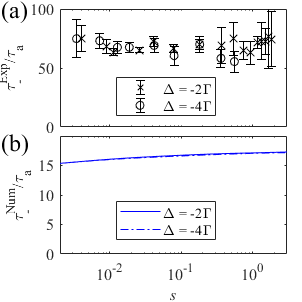} 
\includegraphics[width=0.68\textwidth]{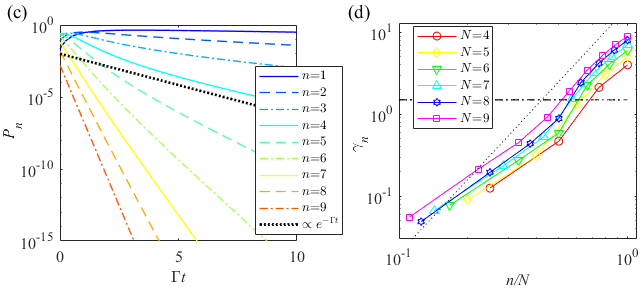}
\caption{(a-b) Lifetime of the long-lived modes as a function of the saturation parameter, from (a) the experimental data and (b) the simulations of the truncated dynamics [same parameters and same conventions for the error bars as in Fig.\,\ref{fig:PopuI}(a)]. (c) Dynamics of the population $P_n$ of the $n$-excitation states for a cloud of $N=9$ atoms, initially driven to steady-state by a strong resonant field before the pump is switched off (the ground state one, $P_0=1-\sum_{n\geq1}$, is not shown here).  (d) Decay rate $\gamma_n$ of the populations $P_n$, computed for $t\in [5; 10]/\Gamma$, as a function of the relative excitation number $n/N$, and for particle numbers from $N=4$ to $9$. The horizontal line marks the $\gamma_n=1$  transition from subradiant to superradiant states, while the dotted black line, which scales as $n^3$, is only a guide for the eye.\label{fig:Pn}}
\end{figure*}

We have benchmarked the truncated method by comparing the late-time, far-field radiated intensity $I_\mathbf{k}\propto \sum_{m,n}e^{i\mathbf{k}.(\mathbf{r}_m-\mathbf{r}_n)}\langle\hat{\sigma}_m^+\hat{\sigma}_n^-\rangle$ to that from exact simulations~\cite{Johansson2012,Johansson2013} [see example in Fig.\,\ref{fig:PopuI}(b)], obtaining accurate results for atomic densities up to $\rho_a\approx 0.03k^3$ (for random distributions with a minimal distance $\rho_a^{-1/3}/2$). Interestingly, we observe that semi-classical simulations (i.e., a truncation at the first order) fail to capture the long-lived states, as they exhibit a single-atom decay dynamics for a strong drive [see Fig.\,\ref{fig:PopuI}(b)]. This is in contrast with the superradiant cascade, known to be described by a semiclassical approach~\cite{Dicke1954,Arecchi1972,MacGillivray1976,Gross1982,Maximo2020}, and it strongly suggests that the subradiant states here studied might be a source of quantum correlations~\cite{Tanas2004}.

The normalized population obtained from simulations with the truncated scheme (using $N=100$ particles) are presented as continuous and dashed lines in Fig.\,\ref{fig:PopuI}(a), and present a good qualitative agreement with the experimental data. The absence of dependence on the detuning validates the earlier hypothesis of negligible frequency shifts.
Although there is some discrepancy in the values of the normalized population (which might be due to the vastly different parameters for the atom number and size of the sample), the observed scaling $P_-\propto s^{1.49}$ is well consistent with the simulated one.

Interestingly, it also shows that even with a saturation parameter as low as $s\approx 3\times10^{-3}$, the linear-optics regime describing the single-excitation physics is not reached, in presence of cooperative effects. The simulations at $b_0=5$ present a nonlinearity threshold at $s_\mathrm{LO}\approx 2\times10^{-3}$, yet it was not possible to obtain a scaling of $s_\mathrm{LO}$ from the low-$b_0$ simulations. Nevertheless, we note that the $N=2$ case discussed earlier suggests that longer-lived states (achieved for smaller distances $kr$) couple less to the pump, which in turn results to lower values of $s_\mathrm{LO}$ [$s_\mathrm{LO} \sim 10^{-4}$ in Fig.\,\ref{fig:scheme}(c)]. Furthermore, it was recently suggested that such a threshold may scale as $\Gamma_n^{2.5}$, with $\Gamma_n$ the $n$-th mode linewidth~\cite{Williamson2020a}: Assuming the subradiant states present linewidths scaling as $\Gamma_n\sim\Gamma/b_0$~\cite{Guerin2016}, saturation parameters orders of magnitude smaller may be necessary to experimentally reach the linear-optics regime for long-lived states.

Increasing the pump power opens the possibility of exploring a much broader part of the $2^N$-dimensional Hilbert space, and thus potentially access much longer lifetimes -- the $N=2$ case only yields one superradiant and one long-lived states. Nonetheless, as observed in Fig.~\ref{fig:Pn}(a), the lifetime of long-lived modes is only marginally affected by the strength of the drive. While the simulations of the truncated dynamics present an increase of $\sim15\%$ in lifetime as the saturation parameter is increased, the experimental error bars do not permit us to identify this increase. 

To understand better the preservation of the long lifetimes, studying all the collective modes of the system is not relevant since the question is rather about which ones are populated by the pump \cite{Guerin2017}. Thus, we monitor the decay dynamics of the population of the $n$-excitation states, $P_n = \mathrm{Tr}(\hat{P}_n\hat\rho)$, where we have introduced the projector
$\hat{P}_n=\sum_{\mathcal{P}}\otimes_{j=1}^n \dyad{e_j}{e_j}\otimes_{m=n+1}^N\dyad{g_m}{g_m}$,
with $\mathcal{P}$ the ${N!}/{n! (N-n)!}$ permutations of a set with $n$ excited atoms and $(N-n)$ in the ground state. 

The evolution of the population of $n$-excitation states is presented in Fig.\,\ref{fig:Pn}(c), where we observe that, after a short transient of order $1/\Gamma$, the lower the excitation number $n$, the slower the decay: Highly excited states decay quickly into low-excitation ones, where the excitations remain for long times, as compared to $1/\Gamma$. A systematic analysis of the dynamics, for atom numbers ranging from $4$ to $9$, reveals that states from the upper part of the Dicke space ($n>N/2$) decay at superradiant rates (even at late times), whereas the lower part ($n\leq N/2$) is characterized by long lifetimes (at late times), see Fig.\,\ref{fig:Pn}(d). Consequently, the short-time (superradiant) emission is realized by highly excited states, whereas the late-time emission comes from low-excitation ones. This is illustrated in Fig.~\ref{fig:PopuI}(b), where the contribution to the radiated intensity of low- and high-excitation states were computed separately. Another consequence is that an increased pump strength populates subradiant states with shorter lifetimes, so the radiation from low-excitation, longer-lived, states may dominate later, see \cite{SuppMat}. Finally, we have checked that starting from a fully inverted system, as originally studied by Dicke~\cite{Dicke1954}, the initial (superradiant) dynamics differs from that of a cloud driven to steady-state with a strong pump, yet the same long lifetimes are eventually observed (not shown here): The superradiant cascade occurs, yet not without some decay toward longer-lived modes, as illustrated in Fig.\,\ref{fig:scheme}(a).

Hence, the $n\geq2$ part of the Hilbert space, which comprises everything beyond linear optics, does not appear to offer access to longer lifetimes, nor does it provide a path to escape long-lived modes. Remarkably, a study on one-dimensional regular chains has reached the similar conclusion that the {\it longest-lived} higher-excitation states present a decay rate that scales as $(n/N)^3$~\cite{Asenjo2017}, in excellent agreement with our simulations [see the dotted line in Fig.\,\ref{fig:Pn}(d)]. This suggests that this feature may be quite universal, beyond the details of geometry and dimensionality of the system.

In conclusion, we have reported on the experimental observation that long-lived subradiant states can have their population enhanced by increasing the pump strength, through a mechanism similar to optical pumping via superradiant states. Surprisingly, the single-excitation (linear-optics) subspace actually presents the longest lifetimes, in contrast with the superradiant states, which reach their fastest rate at half-excited population.

The faster decay of higher-excitation states leads to the puzzling question of the entanglement of these long-lived states created through decay processes~\cite{Tanas2004}, as the difference between a true single-excitation state \cite{Scully2006,Frowis2017} and its linear-optics separable counterpart resides in the multi-excitation component \cite{Eberly2006,Bienaime2011}. The failure of a semi-classical approach to describe properly those decay processes is a further argument to support the idea that the subradiant states might be appropriate to store entanglement or quantum correlations. Finally, the present scheme could in principle be applied to cavity setups, where the light from subradiant states could be collected more efficiently in the cavity modes, for example using external fields~\cite{Facchinetti2016}.

 \begin{acknowledgments}
 Part of this work was performed in the framework of the European Training Network ColOpt, which is funded by the European Union (EU) Horizon 2020 program under the Marie Sklodowska-Curie action, grant agreement No.\,721465, and the project ANDLICA, ERC Advanced grant No. 832219. We also acknowledge funding from the French National Research Agency (projects PACE-IN ANR19-QUAN-003-01 and QuaCor ANR19-CE47-0014-01). R.\,B., T.\,S.\,E.\,S., and C.\,J.\,V.\,-B. benefited from Grants from S\~ao Paulo Research Foundation (FAPESP, Grants Nos. 2018/01447-2, 2018/15554-5, 2018/12653-2, 2019/13143-0, 2019/02071-9, and 2019/11999-5) and from the National Council for Scientific and Technological Development (CNPq, Grant Nos.\,302981/2017-9, 409946/2018-4, and 307077/2018-7). R.\,B. and R.\,K. received support from the project CAPES-COFECUB (Ph879-17/CAPES 88887.130197/2017-01). P.\,W. received support from the Deutsche Forschungsgemeinschaft (Grant No.\,WE 6356/1-1).
 \end{acknowledgments}

 \textbf{Contributions} -- A.C. and P.W acquired the experimental data. A.C. and W.G performed systematic analysis of the experimental data. T.S.E.S. contributed to the development of theoretical and numerical tools and, together with N.A.M., realized the numerical simulations. C.J..V-.B., together with R.B., contributed with the proposal of the conceptual work and analytical results. R.K., W.G. and R.B. supervised the project. All authors worked on the interpretation of the data and contributed to the final manuscript.


\bibliography{BiblioCollectiveScattering}{}

\appendix

\section*{Subradiance with saturated atoms: population enhancement of the long-lived states:
Supplemental material}

\section{Coupled dipole model for $N$ two-level atoms in free space}

The atomic cloud is modelled as an ensemble of $N$ two-level emitters with positions $\mathbf{r}_j$ and rising/lowering operators $\hat\sigma_j^{+}/\hat\sigma_j^{-}$ between their ground and excited states $g$ and $e$; $\omega_a=kc=2\pi c/\lambda$ is the transition frequency and $\Gamma$ its linewidth. The cloud is driven by a near-resonance monochromatic field with Rabi frequency $\Omega(\mathbf{r})$, detuned from the transition by $\Delta$. Within the Markov and rotating-wave approximations, the coupled dynamics of the evolution of the density matrix $\hat\rho$ describing the atomic dipoles is obtained from the master equation $\dot{\hat\rho}=-i [H, \hat\rho] + \mathcal{L}(\hat\rho)$, where the coherent Hamiltonian $\hat H$ and dissipative dynamics $\mathcal{L}$ are given, in the pump frame, by~\cite{Stephen1964, Lehmberg1970, Friedberg1973}:
\begin{subequations}
\begin{align}
\hat H &= - \Delta \sum_m  \hat \sigma_m^+  \hat \sigma_m^- + \frac{1}{2} \sum_m \left[ \Omega(\mathbf{r}_m) \hat \sigma_m^+  + h.c. \right] \nonumber \\
& +  \frac{1}{2}\sum_{m, n\neq m} \Delta_{mn} \hat \sigma_m^+ \hat \sigma_n^-, \label{eq:H_dd}
\\ \mathcal{L}(\hat \rho) &= \frac{1}{2}\sum_{m,n} \Gamma_{nm} \left(2\hat \sigma_m^- \hat \rho \hat \sigma_n^+ - \hat \sigma_n^+ \hat \sigma_m^-\hat \rho- \hat\rho \hat \sigma_n^+ \hat \sigma_m^- \right).\label{eq:diss}
\end{align}
\end{subequations}
The diagonal term corresponds to the single-atom dynamics, $\Gamma_{nn}=\Gamma=1$ and $\Delta_{nn}=0$, while the coupling terms are given by $\Delta_{mn}=-\Gamma\cos(kr_{mn})/(kr_{mn})$ and $\Gamma_{mn}=\Gamma\sin(kr_{mn})/(kr_{mn})$,
with $r_{mn}=|\mathbf{r}_m-\mathbf{r}_n|$. This model corresponds to a `scalar light' approximation: although polarization can play an important role in subwavelength clouds~\cite{Gross1982,Cremer2020}, the experimental situation is that of a dilute cloud (atomic density $\rho_a\approx 0.06\lambda^{-3}$), where a scalar description of the light is a good approximation~\cite{Cipris2020}.

\section{Calibration of the saturation parameter} 

In this experiment, it is important to have a proper calibration of the saturation parameter, defined as
\begin{align}
    s(\delta)=g\dfrac{s_0}{1+4(\Delta/\Gamma)^2}, \label{satpar}
\end{align}
where $g=7/15$ is the degeneracy factor of the $\ket{F=2} \rightarrow \ket{F'=3}$ D$_2$ transition of $^{87}$Rb for a statistical mixture of equally populated Zeeman sublevels and $s_0=I/I_{\text{sat}}$ is the on-resonance saturation parameter, with $I_{\text{sat}}=1.6\, \text{mW}/\text{cm}^2$ the  saturation intensity. A first evaluation of the saturation parameter at the center of the beam can be obtained from the measurement of the beam power and its waist ($1/e^2$ radius $w=5.3$ mm), but this is usually not precise because of a number of effects: losses along the beam path, beam not perfectly Gaussian, atomic cloud not perfectly at the center of the beam, etc. A calibration method based on the interaction with the atoms is thus preferable. Hereafter we use the label $s'(\Delta)$ for the saturation parameter that is determined by measuring the power and waist of the probe beam, while for the properly calibrated saturation parameter we use $s(\Delta)$. Note that the probe power is measured simultaneously to the data acquisition by a dedicated detector.

The first calibration method is based on the measurement of the fluorescence level. The atomic cloud ($b_0=37\pm1$) was illuminated by the probe beam with $\Delta=-4\Gamma$ and the fluorescence signal was recorded as a function of $s'(\Delta)$ 
. Since the total scattering rate is $\propto s/(s+1)$, we fitted the measured fluorescence level by $f=Bs'/(Cs'+1)$ and we obtained a correction factor for the saturation parameter $C=0.36\pm0.04$.


The other two calibration methods rely on hyperfine depumping into the $\ket{F=1}$ ground state. Although the transition of interest in this experiment is $\ket{F=2}\rightarrow \ket{F'=3}$, when the probe beam is largely detuned to the red from that transition, there is a significant probability of exciting the $\ket{F'=2}$ state, from which atoms can decay into the dark $\ket{F=1}$ state. The corresponding depumping rate is given by
\begin{equation*}
\Gamma_{\text{depump}}=p_{21}\dfrac{\Gamma}{2}\dfrac{s_{22}}{s_{22}+1}, \,\, \mathrm{with} \,\,
s_{22}=g_{22}\dfrac{s_0}{1+4(\Delta_{22}/\Gamma)^2},
\end{equation*}
where $p_{21}=1/2$ is the decay probability of $\ket{F'=2}$ towards $\ket{F=1}$, $s_{22}$ is the saturation parameter for the $\ket{F=2}\rightarrow \ket{F'=2}$ transition, with $g_{22}=1/6$ the corresponding degeneracy factor, and $\Delta_{22}=44\Gamma+\Delta$ the corresponding detuning. 
By measuring the number of atoms in the $\ket{F=2}$ as a function of time, $N(t)=N_0\text{exp}(-\Gamma_{\text{depump}}t)$, we can extract the depumping rate and then the saturation parameter.

At sufficiently large detuning, the fluorescence level is proportional to the number of atoms, so the depumping rate can be obtained by measuring the fluorescence level as a function of the laser duration. In fact, since we illuminated the atomic cloud by series of 12 pulses of duration $10\,\mu$s each, we added up the duration of subsequent pulses. By fitting an exponential decay to the measured fluorescence level 
, we obtained the depumping rate, from which we determined the saturation parameter. 
This method yielded the correction factor $C=0.32\pm0.01$.

Another way of obtaining the depumping rate is by measuring the optical thickness as a function of the laser duration with and without a repumping stage just before absorption imaging. Then, by fitting the ratio by an exponential decay 
, we extracted the depumping rate and with that we obtained the correction factor $C=0.34\pm0.01$. 

The three methods are in very good agreement and have similar uncertainties. Therefore, we use the average of the three methods: $C=0.34\pm 0.02$.

\section{Measurement of optical thickness and temperature}

In this experiment we probe the atomic sample with a varying saturation parameter, up to relatively large values. The interaction with the light can thus have significant effects on the atomic cloud, in particular heating and pushing. It is therefore important to characterize the atomic sample taking into account those effects.

To this end, the optical depth of the cloud and its temperature are measured simultaneously to the data acquisition using interlaced cycles of subradiance measurements and absorption imaging as described in \cite{Weiss2019}. The same probe beam is used for both but absorption imaging is always performed at low saturation parameter and large detuning $\Delta=-4\Gamma$.

More precisely, absorption imaging was performed instead of the fluorescence measurement once every 250 cycles. Since during the data acquisition we are probing the atomic cloud with series of 12 laser pulses, to measure the optical thickness of the cloud corresponding to the $n$th pulse, we apply $(n-1)$ probe pulses before absorption imaging at the time corresponding to the $n$th pulse. 
This protocol enables us to have a good calibration of $b_0$ while probing our sample with laser pulses of different intensities.

The absorption imaging was also done for different times of flight of ballistic expansion, without and with a few applied probe pulses before, in order to measure the initial temperature of the cloud, as well as the heating induced by the pulses.

From those measurements we were able to extrapolate the temperature of the cloud after each of the 12 applied pulses. While the minimal temperature of the cloud that we measured is $T\approx100\,\mu$K, the maximum temperature, considering the heating, is $T\approx700\,\mu$K. As shown in \cite{Weiss2019}, in this range of temperature, subradiance is not significantly affected: only a very slight decrease of the subradiant lifetime can be expected for the maximum considered temperature. 

Another effect that could be relevant is the radiation-pressure force exerted by the probe beam on the cloud, inducing a velocity along the beam direction and correspondingly a Doppler shift, which changes the detuning seen by the atoms and thus the effective saturation parameter. This pushing effect can easily be computed and we have checked that, even for the highest saturation parameter, it only induces a very small reduction of the saturation parameter, which does not affect our results. This is confirmed by the fact that different values of $b_0$, corresponding to different times of flight and thus different numbers of applied pulses, yield similar results~\cite{SuppMat}.

Finally, the expansion of the cloud during the pulse series is also responsible for a slight decrease of the effective intensity interacting with the cloud (due to the finite beam waist), of at most $\sim 20 \%$. It does not affect significantly any of the presented results.

\section{Subradiant decay curves}

\begin{figure}[b]
\centering
\includegraphics[width=\columnwidth]{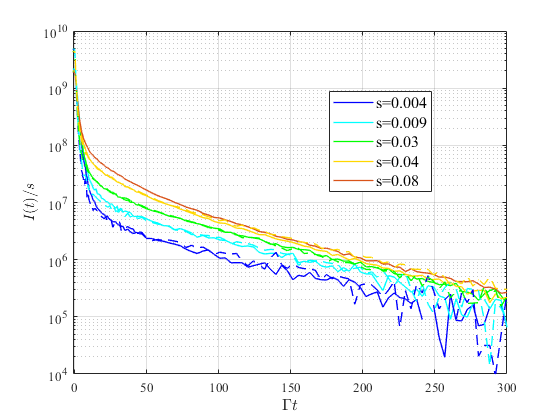} 
\caption{Temporal dynamics of the scattered intensity at the switch-off of the driving field, for different saturation parameters $s$ as well as for $\Delta=-2\Gamma$ (solid line) and $\Delta=-4\Gamma$ (dashed line). The intensity curves are normalized by $s$. The amplitude of the slow decay increases with $s$ and it is similar for both detunings.} \label{fig:decaycurve50}
\end{figure}

In Fig.\,S\ref{fig:decaycurve50} we show typical experimental subradiant decay curves. This figure illustrates well the influence of the saturation parameter and detuning on the population of long-lived modes. Here, the scattered light intensity is normalized by the saturation parameter, and we only show values for $s \ll 1$. We observe that the amplitude of the slow decay, even after this normalization, increases with the saturation parameter, which illustrates well the superlinear behavior of the long-lived mode populations.

\section{Superlinear growth with $s(\Delta)$ for several $b_0$}

\begin{figure}[h]
\centering
\includegraphics[width=\columnwidth]{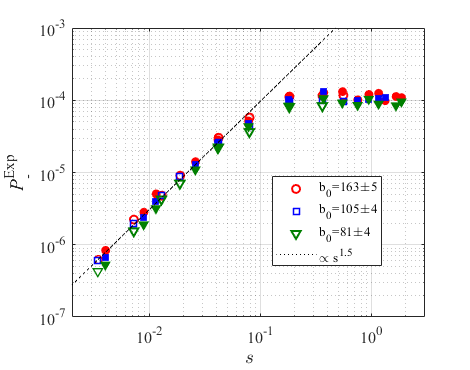} 
\caption{Normalized population $P_-$ of long-lived modes as a function of the saturation parameter $s$ for three different $b_0$, as well as $\Delta=-2\Gamma$ (filled circles) and $\Delta=-4\Gamma$ (empty circles). The results were obtained using the fit window $\Gamma t\in [150,250]$.} \label{fig:different_b0}
\end{figure}

In the main text the superlinear growth of the long-lived mode population as a function of the saturation parameter has been shown for an atomic sample of resonant optical thickness $b_0=54 \pm 2$ and two different detunings [Fig.\,2(a)]. Here we demonstrate that this observation is robust by showing data acquired for different values of $b_0$. In Fig.\,S\ref{fig:different_b0} we show the normalized population $P_-$ (see main text) as a function of $s$ for $b_0 = 81 \pm 4$, $b_0 = 105 \pm 4$ and $b_0 = 163 \pm 5$. The obtained exponents of the power-law fit at low $s$ are $\beta \approx 1.47$, $\beta \approx 1.46$ and $\beta \approx 1.49$, respectively, demonstrating that the power-law scaling $\beta \sim 1.5$ is independent of $b_0$.

\section{Population of modes of different lifetimes\label{App:Lifetimes}}


Here we have investigated the behaviour of the subradiant modes with different lifetimes by monitoring the emission decay over different time windows: Later time analyses allow us to study longer-lived modes as they take longer to deplete. Over all the selected time windows, the super-linear scaling of the subradiant population $P_-$ with the saturation parameter $s$ is observed, see Fig.\,S\ref{fig:diff_fit}(a). The study of the associated lifetimes in Fig.\,S\ref{fig:diff_fit}(b) confirms that later-time dynamics is dominated by more subradiant modes. One can also observe that only the lifetimes extracted from the most-delayed fit window are independent of $s$ because they correspond to single-excitation collective states. Indeed, increasing the pump strength leads to exploring subradiant states with shorter lifetimes, so the radiation from low-excitation subradiant states, with the largest lifetimes, may dominate only at later times.

Interestingly, in the regime where the populations grow super-linearly, they all have the same values (the curves of Fig.\,S\label{fig:diff_fit}(a) collapse together). This suggests that the same population can be reached independently of the considered lifetimes. A first hint to understand the underlying mechanism can be found again in the case of $N=2$ atoms: There, both the transitions $\ket{ee}\to\ket{-}$ and $\ket{-}\to\ket{gg}$ present a decay rate $\Gamma_-$. In a simple rate-equation approach, the population stored by optical pumping is expected to depend on the ratio between these rates, therefore the explicit dependence on $\Gamma_-$ disappears.

From a numerical perspective, the low-$b_0$ clouds that can be simulated do not allow us to explore the behavior of modes with different lifetimes. As an alternative, we consider a system of $N=5$ atoms in a small volume, organized as a regular chain for simplicity. Thus, the single-excitation collective modes are composed of one superradiant mode and four subradiant ones. Monitoring the steady-state population of these modes as a function of the saturation parameter reveals that, despite they exhibit very different levels in the ``linear-optics regime''
($P_-\propto s$), the three most subradiant modes acquire the same populations in the regime where the modes are optically pumped [Fig.\,\ref{fig:diff_fit}(c)]. Different chain lengths presented some fluctuations, but generally the populations have a strong tendency to reach the same values in that regime, suggesting that optical pumping induces an even population of the long-lived modes independently of their lifetimes.

\begin{figure*}
\centering
\includegraphics[width=\textwidth]{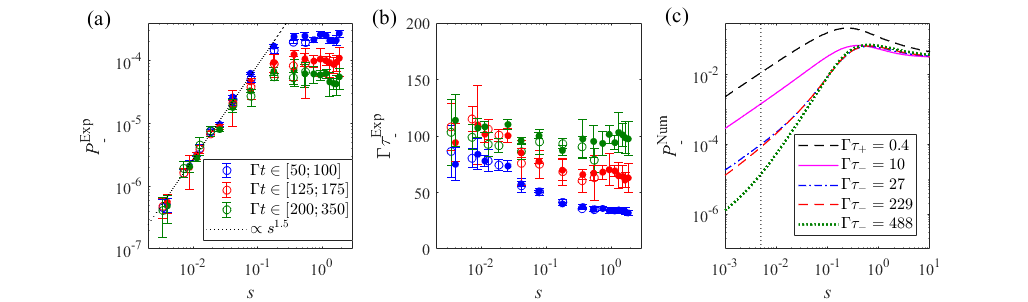} 
\caption{(a) Subradiant population $P_-$ (b) and associated lifetimes $\Gamma\tau_-$ as a function of the saturation parameter, obtained from the fit over different time windows. The optical depth is $b_0\approx 80$. Filled and empty symbols are for $\Delta=-2 \Gamma$ and $\Delta=-4 \Gamma$, respectively. (c) Population of the single-excitation collective modes for a linear regular chain of length $kL=1$, with a pump in the direction of the chain and a detuning $\Delta=-200\Gamma$. The vertical black dotted line marks the saturation value around which the population of the most subradiant modes departs from the linear-optics behaviour.}\label{fig:diff_fit}
\end{figure*}

The fact that the subradiant populations saturate, for large values of $s$, at different values in the experiment [Fig.\,S\ref{fig:diff_fit}(a)], is not observed in our small-system simulations [Fig.\,S\ref{fig:diff_fit}(c)]. It may thus be related to the large experimental cloud possessing a huge spectrum of subradiant modes, with a strong degeneracy of the energies (that is, a strong overlap between various modes), different from the latter test-system. Unfortunately, the simulations with the truncated system, limited to $b_0\approx 5$, do not allow us to investigate the different lifetimes using larger systems ($N=100$). One can only conclude that the huge Hilbert space at stake still holds many surprises. 

\ \\

\onecolumngrid

\end{document}